\newcommand{\bea}{\begin{eqnarray}}
\newcommand{\eea}{\end{eqnarray}}
\newcommand{\nn}{\nonumber}
\newcommand{\Ga}{\Gamma}
\newcommand{\g}{\gamma}

\newcommand{\s}{\sigma}

\newcommand{\bt}[1]{{\bar t}}

\documentclass[twocolumn,showpacs,preprintnumbers,amsmath,amssymb]{revtex4}

\usepackage{graphicx}
\usepackage{dcolumn}
\usepackage{bm}

\begin{document}

\preprint{}

\title{Role of Bell Singlet State in the Suppression of Disentanglement

}

\author{Ru-Fen Liu}
 \email{fmliu@phys.ncku.edu.tw}
\author{Chia-Chu Chen}%
 \email{chiachu@phys.ncku.edu.tw}
\affiliation{%
National Cheng-Kung University, Physics Department, 70101, 1
University Road, Tainan, Taiwan, R. O. C. }%

\begin{abstract} The stability of entanglement of two atoms in a cavity
is analyzed in this work. By studying the general Werner states we
clarify the role of Bell-singlet state in the problem of
suppression of disentanglement due to spontaneous emission. It is
also shown explicitly that the final amount of entanglement
depends on the initial ingredients of the Bell-singlet state.

\end{abstract}

\maketitle

One of the specific features of quantum world is the existence of
quantum coherence which forms the basis of describing wide
varieties of phenomena including superconductivity and
Bose-Einstein condensation of cold atoms. During the last decade,
another aspect of quantum coherence, namely, quantum
entanglement\cite{EPR}, has been recognized as the essential
element of quantum computing\cite{Issac}. In order to realize
quantum information processing, stability of entanglement of
quantum subsystems is one of the important problems that requires
careful analysis. Instabilities of quantum entanglement can be
generated through different mechanisms\cite{CDec}. In general, an
entangled state of a closed system can be disentangled by its own
dynamics\cite{FC}. On the other hand, due to decoherence, system
and environment interaction might not preserve initially entangled
state. However, decoherence can also be a dynamical effect if one
includes the quantum fluctuation of vacuum. In fact, such
fluctuation is the origin of spontaneous emission which can reduce
entangled state to separable state via photon emission.

The recent work of Yu and Eberly\cite{Eber} has discussed the
finite-time disentanglement via spontaneous emission. In their
system two non-interacting atoms are coupled to two separate
cavities(environments). As a result, the dynamical evolution of
the atoms are independent and, depending on initial state, the
effect of spontaneous emission can drive the system to
disentangled in finite time. However it is not clear if the
disentangle phenomenon will persist if the atoms are allowed to
interact. Intuitively, it is easy to imagine that for two atoms
interacting in a lossless cavity, the photon emits by one atom
during spontaneous emission can be absorbed by the other. As a
result, entanglement might be preserved through this mechanism. In
fact the above photon process is equivalent to the interaction
between atoms by exchanging photon. Furthermore it is also more
practical for constructing the quantum circuit inside one cavity
instead of distribute the atoms in different separate ones.
Consequently, it is inevitable to include the effects of
interaction among atoms for any discussions on disentanglement via
spontaneous emission. This problem has also been addressed in the
interesting work by Tana\'{s} and Ficek\cite{Tanas}. By putting
two atoms inside the same cavity they showed that the entanglement
exhibits oscillatory behavior, and the amount of entanglement is
directly related to the population of the slowly decaying
Bell-singlet state in the long time limit. Their results is
interesting since it indicates that the Bell-singlet state is
stable against photon emissions. To justify this point more
concretely, it is necessary to explore further on the role of
Bell-singlet in the suppression of disentanglement. This is what
will be discussed in this work.

Our  model system is the same as the one employed in \cite{Tanas}
except for the fact that we neglect the spatial dependent of the
atoms to avoid complication. Such simplification will make the
role of spin singlet state more prominent in the suppression
mechanism. The model consists of two two-level atoms $A$ and $B$
inside a lossless cavity. These atoms are considered as identical
and allowed to interact by exchanging photon inside the cavity
which is viewed as the environment. The coupling between the
system and the environment is the origin of the disentanglement.
The Hamiltonian of the total system is given by
$H_T=H_s+H_{sb}+H_b$. $H_s$, $H_b$ and $H_{sb}$ are atomic, the
bath and atoms-bath interaction Hamiltonian
respectively($\hbar=1$): \bea H_s &=& \frac{1}{2}\omega_0\Sigma_z
\\ H_b &=& \sum_k \omega_k (a^{\dagger}_ka_k+\frac{1}{2}) \\
H_{sb} &=& \sum_k (g_k^*\Sigma_-a_k^{\dagger}+g_k\Sigma_+a_k)\eea
where $\Sigma_i\equiv\s_i^A+\s_i^B$, $i$ can either be $\{x, y,
z\}$ or $\{+, -\}$ for raising and lowering operations and
$a_k(a_k^{\dagger})$ is the photon annihilation(creation) operator
of mode $k$. Here we have assumed that these atoms are identical
such that $\omega_A=\omega_B\equiv \omega_0$ and they couple to
photon mode $k$ with the same strength $g_k$. In this work we
assume that the atoms are entangled but not with the bath at
$t=0$. Furthermore, the initial bath state is assumed to be the
vacuum state. The initial total state is then given by the
following product state, \bea|\psi_{total}\rangle =
|\psi\rangle_{AB}\otimes|0\rangle.\eea Here $|\psi\rangle_{AB}$ is
the entangled initial state of the atoms and $|0\rangle$ denotes
the vacuum state of the cavity. The master equation of atoms in
the Schr\"{o}dinger picture can be obtained as follows: \bea
\dot{\rho}_t&=&-i[H_s,\rho_t]-\{\Sigma_+f(t)\Sigma_-\rho_t-
f(t)\Sigma_-\rho_t\Sigma_+ \nn
\\ &&+\rho_tf^{\dagger}(t)\Sigma_+\Sigma_--
f^{\dagger}(t)\Sigma_+\rho_t\Sigma_-\}. \eea Here
$f(t)\equiv\sum_k \int_0^t dt' C_k(t-t')e^{i\omega_0(t-t')}$ with
$C_k(t-t')$ given by the photon correlation function
 \bea
C_k(t-t')=Tr(\tilde{A}_k(t)\tilde{A}_k^{\dagger}(t')
|0\rangle\langle 0|)\eea where $\tilde{A}_k(t)\equiv\sum_k g_k
\tilde{a}_k(t)$ and $\tilde{a}_k(t)$ is the photon annihilation
operator in interaction picture. With $f(t)=f_R(t)+if_I(t)$, one
can further arrange Eq.(5) into unitary and decoherent evolutions
as follows: \bea \dot{\rho}_t&=&-i[H_s+f_I(t)\Sigma_+
\Sigma_-,\rho_t]\nn
\\ &&-f_R(t)\{[\Sigma_+,\Sigma_-\rho_t]+ [\rho_t\Sigma_+,\Sigma_-] \}.
\eea The physical meaning of Eq.(7) can be understand by expanding
the commutators. Inside the first commutator, the hamiltonian
which contributes to the unitary evolution contains the energy
eigenvalues to the second order corrections and dipole
interactions between $A$ and $B$ atoms. One can see that the
coupling of dipole interaction is identical to the energy
correction. This is due to the fact that both atoms couple to
photon with the same strength. The rest of Eq.(7) which is
non-unitary could intuitively imply decoherent evolution. However
it turns out that in our case this intuitive picture is illusive.
Apart from the spontaneous real photon emission process which
definitely gives rise to decoherence, the non-unitary dynamics
also includes more non-local photon-exchange interactions which
turns out to be the main driving force for the suppression of
disentanglement mentioned earlier. Explicitly, the master equation
is: \bea\dot{\rho}_t&=&
-i[\frac{1}{2}(\omega_0+f_I(t))\Sigma_z+f_I(t)
(\s_+^A\s_-^B+\s_+^B\s_-^A), \rho_t] \nn \\ && -f_R(t)\{
\s_+^A\s_-^B\rho_t+\rho_t\s_+^A\s_-^B+\s_+^B\s_-^A\rho_t+\rho_t\s_+^B\s_-^A
\nn \\ && \ \ \ \ \ \ \ \ \ \ \
-2\s_-^A\rho_t\s_+^B-2\s_-^B\rho_t\s_+^A \} \nn
\\ &&-2f_R(t)\{\s_+^A\s_-^A\rho_t+\rho_t\s_+^A\s_-^A
-\s_-^A\rho_t\s_+^A-\s_-^A\rho_t\s_+^A\} \nn \\
&&-2f_R(t)\{\s_+^B\s_-^B\rho_t+\rho_t\s_+^B\s_-^B
-\s_-^B\rho_t\s_+^B-\s_-^B\rho_t\s_+^B\}. \nn \eea The general
solution for the master equation can be found by constructing the
Kraus operator $K_\mu(t)$ which gives the density matrix $\rho(t)$
in terms of the initial state $\rho(0)$ as \bea\rho(t)=\sum_{\mu}
K_\mu(t)\rho(0)K_{\mu}^{\dagger}(t),\eea where the Kraus operators
$K_\mu(t)$ satisfy $\sum_{\mu} K_\mu(t)K_{\mu}^{\dagger}(t)=I$ for
all $t$. The advantage of using the Kraus representation is the
fact that $\rho(t)$ satisfies all the requirements of the density
matrix which are positivity and $Tr\rho=1$. However, in this work
it is more easier to obtain $\rho(t)$ by solving the coupled
differential equation directly. Due to the non-unitary evolution
of the master equation, such $\rho(t)$ does not a priori satisfy
the above requirements of density matrix and therefore it requires
extra care to ensure the results are satisfactory. Accordingly, we
have checked in all calculations that our solution are indeed a
good density matrix for all time.

It is well-known that a good definition of entangled mixed state
is still lacking for general system. However, for $2\times 2$
system it has been shown by Peres\cite{Peres} and
Horedicki's\cite{Hore} that Positive Partial Transpose (PPT) of
the density matrix is a good criterion for characterizing the
separability of states. It will be shown latter that both
definitions are equivalent. Furthermore, the distillation
protocols first invented by Bennett et. al.\cite{Ben} are
performed by LOCC and can convert a large number of mixed state
with ingredient of entanglement into a smaller number of maximally
entangled pure state which is the most important resource in
quantum information processing (QIP). There are many efforts on
figuring out the relation between distillability and separability
of mixed state density matrix\cite{Horo2}. For $2\times 2$ system,
it has been shown that the mixed state can be distilled if it
violates PPT criterion\cite{Horo2}. In this work, we will shown
latter that the state of two coupled atoms in cavity undergoing
the dynamical bath interaction become inseparable, stable and
distillable.

In what follows we will choose the initial state with the form in
computational basis of $\{|++\rangle, |+-\rangle, |-+\rangle,
|--\rangle\}$ where $|-\rangle$($|+\rangle$) represents the
ground(excited) state of the atom: \bea
\rho=\left(\begin{array}{cccc}
\rho_{11}&0&0&0\\0&\rho_{22}&\rho_{23}&0\\0&\rho_{32}&\rho_{33}&0\\
0&0&0&\rho_{44}\end{array}\right),\eea these states contains a
special subclass of mixed state, the general Werner states
$W_F$\cite{Werner,Ben2} parameterized by a single real parameter
$F$ is: $\rho_{11}=\rho_{44}=\frac{1}{3} (1-F)$,
$\rho_{22}=\rho_{33}=\frac{1}{6}(1+2F)$ and
$\rho_{23}=\rho_{32}=\frac{1}{6}(1-4F)$. One can easily check that
the conditions of density state require that $0\leq F\leq1$ and
$W_F$ is inseparable for $F>\frac{1}{2}$. $F$ which can be regards
as the fidelity of $W_F$ , $\langle\Psi^-|W_F|\Psi^-\rangle$,
relative to singlet state also quantifies the upper bound of the
distillable maximally entangled singlet states by LOCC. We note
that Wootter's concurrence\cite{woo} is given by $C(\rho)=
Max\{0,\lambda_1-\lambda_2-\lambda_3-\lambda_4\}$ where
$\lambda$'s are the square root of the eigenvalues of spin flow
matrix defined by $\rho$: $R=\rho(\s_y^A\otimes\s_y^B)\rho^*
(\s_y^A\otimes\s_y^B)$, subtracting in decreasing order. A state
contains no entanglement with $C=0$, while maximally entangled
with $C=1$. Hence, the concurrence of $\rho$ is given by \bea
C(\rho)=2Max\{0,|\rho_{23}(t)|-\sqrt{\rho_{11}(t)\rho_{44}(t)}\}.
\eea As to the PPT criterion, \bea\rho_{11}(t) \rho_{44}(t)
\geq|\rho_{23}(t)|^2,\eea which gives the same consequence of
concurrence in Eq. (10). For convenience of following discussion,
we define $\xi(t)\equiv\rho_{11}(t)\rho_{44}(t)-|\rho_{23}(t)|^2$
which is positive for separability, whereas negative for entangled
state.
\begin{figure}
\includegraphics{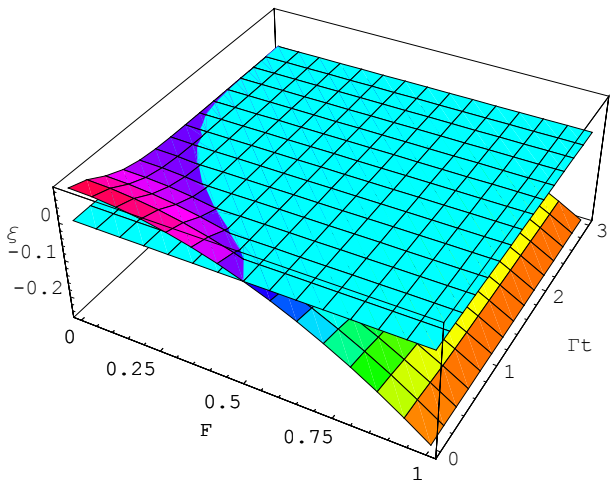}
\caption{\label{fig:epsart}(Color online) The $\xi(t,F)$ plot
which is PPT criterion with $0\leq F\leq 1$. For $F>1/2$, $\xi$ is
negative for all time, which indicates that the inseparability is
preserved under evolution. For $F\leq 1/2$, initially the $\rho_F$
is separable due to positive of $\xi$, while within finite time,
they become inseparable.}
\end{figure}

Due to the special form of the initial states considered in this
work, we will give only the relevant matrix elements explicitly.
In the Markovian limit such that $\int_0^t
dt'f_R(t')\equiv\frac{\Ga}{2} t$ and $\int_0^t
dt'f_I(t')\equiv\frac{\g}{2} t$,these matrix elements are:
\begin{subequations} \bea \rho_{11}(t)=\rho_{11}(0)e^{-2\Ga
t} \\ \rho_{22}(t) = \frac{S_-+e^{-2\Ga t}S(t)}{4}-e^{-\Ga
t}\rho_I(0)\sin\g t \\ \rho_{33}(t)= \frac{S_-+e^{-2\Ga
t}S(t)}{4}+e^{-\Ga t} \rho_I(0)\sin\g t \\ \rho_{23}(t) =
\frac{-S_-+e^{-2\Ga t}S(t)}{4}+ ie^{-\Ga t} \rho_I(0)\cos\g t
\\ \rho_{32}(t) = \frac{-S_-+e^{-2\Ga t}S(t)}{4}- ie^{-\Ga t}\rho_I(0)\cos\g t
\\ \rho_{44}(t) = 1-\rho_{11}(t)- \rho_{22}(t)-\rho_{33}(t).\eea
\end{subequations} Here, $S(t)\equiv S_++4\rho_{11}(0)\Ga t$ and
$S_-\equiv\rho_{22}(0)+\rho_{33}(0)-\rho_{23}(0)-\rho_{32}(0)$,
$S_+\equiv\rho_{22}(0)+\rho_{33}(0)+\rho_{23}(0)+\rho_{32}(0)$. We
also denote $\rho_I(0)$ as the imaginary part of $\rho_{23}(0)$.
It is obvious that the exponential damping factors in these
equations are due to non-unitary evolution.

\begin{figure*}
\includegraphics{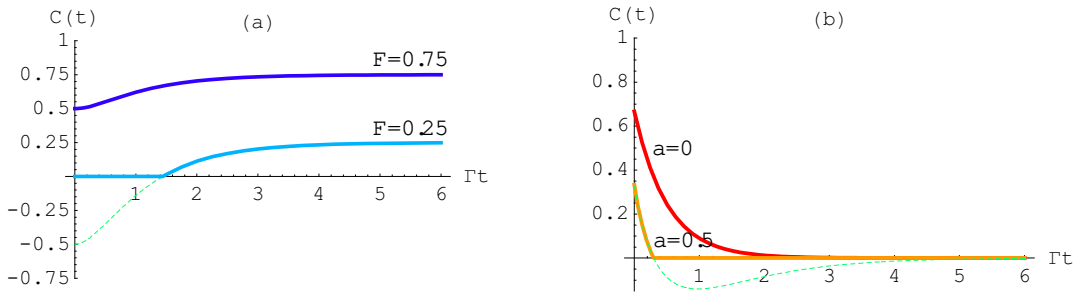}
\caption{\label{fig:wide}(Color online)(a) Concurrence of Werner
state for the cases of $F=0.75$ and $F=0.25$. One can see the
entanglement is enhanced and become stable for both case within
finite time. (b) Concurrence of $\tilde{\rho}(t)$ for the cases of
$a=0.5$ and $a=0$. Disentanglement can always happen suddenly or
asymptotically.}
\end{figure*}

Now we are in the position to show the suppression of
disentanglement in this system. Taking the general Werner states
as initial state, $\rho(0)=W_F\equiv\rho^F(0)$, the evolution is
then given by
\begin{subequations} \bea
\rho_{11}^F(t)= \frac{(1-F)}{3}e^{-2\Ga t} \\
\rho_{22}^F(t)= \frac{F}{2}+e^{-2\Ga t}
\{\frac{(1-F)}{6}+\frac{(1-F)}{3}\Ga t\} \\
\rho_{23}^F(t)= -\frac{F}{2}+e^{-2\Ga t}
\{\frac{(1-F)}{6}+\frac{(1-F)}{3}\Ga t\} \\ \rho_{44}^F(t) = 1-F
-\frac{2(1-F)(1+\Ga t)}{3}e^{-2\Ga t}. \eea
\end{subequations} Here, $\rho_{33}^F(t)=\rho_{22}^F(t)$ and
$\rho_{32}^F(t)=\rho_{23}^F(t)$. Note that the evolution effect
comes solely from the non-unitary evolution. The numerical result
of $\xi(t,F)$ is shown in Fig.(1). For any initially entangled
Werner states with $F>1/2$, one can see from Fig.(1) that
$\xi(t,F)<0$ and as a result the inseparable state remains
entangled for all time. Hence, there is \textbf{NO}
disentanglement effect at all even with spontaneous emission.
Moreover the entanglement can be enhanced and attains stable state
in finite time. To be concrete, we present the results of enhanced
entanglement in Fig.(2a) where we plot the time dependence of
concurrences for two initial states with $F=0.75$ and $F=0.25$.
Our result clearly shows that, for $F=0.75$, the degree of
entanglement is enhanced and reaches a saturated value in finite
time. For the case of $F=0.25$ it is noted that $W_{0.25}$ is just
an equal mixing of all possible eigenstates of the system:
$W_{0.25}=\frac{1}{4}\mathbf{I}$. Accordingly, $W_{0.25}$ is a
classically correlated state with no entanglement. However our
result in Fig.(2a) has shown that entanglement is generated in
finite time.  This last result of generating entangled state from
no-entangled state naturally raises an interesting question,
namely, will it be true that entanglement can always be stabilized
or even enhanced with the presence of non-unitary evolution?
Unfortunately, the answer is negative! To see this, let us
consider the following entangled initial state $\tilde{\rho}$ with
a single parameter $0\leq a\leq 1$:
$\tilde{\rho}_{11}=\frac{1}{3}a$, $\tilde{\rho}_{44}=
\frac{1}{3}(1-a)$ and $\tilde{\rho}_{22}=\tilde{\rho}_{33}=
\tilde{\rho}_{23}=\tilde{\rho}_{32}=\frac{1}{3}$. Note that
$\tilde{\rho}$ is entangled for all $a$ initially($C(0)\neq 0$).
However, for instance, entanglement of the initial state with
$a=0$ is decaying asymptotically, while for $a=0.5$, entanglement
is vanishing abruptly(See Fig.(2b)). This might sound puzzling and
seem contradicting to the result of the classically correlated
state, namely $W_{0.25}$. However, this contradiction is illusive.
To resolve this puzzle, first we note that the Bell-singlet state
has been shown being much more stable than the other Bell states
under dynamical evolution\cite{FC}. Secondly, the initial state
$\tilde{\rho}$ is a mixture of Bell states without the
Bell-singlet. Thus it seems that the instability of the
non-singlet Bell states is probably the reason for disentanglement
to happen. If this is true then the existence of singlet state is
the essential ingredient for enhanced entanglement to happen. To
certify this point, one can add a small amount of singlet states
to $\tilde{\rho}$. Then the modified $\tilde{\rho}^{\epsilon}$ can
be expressed as: \bea
\tilde{\rho}^{\epsilon}&=&\frac{1}{3}\{2\epsilon|\Psi^-
\rangle\langle\Psi^-|+a|++\rangle\langle++|+2|\Psi^+\rangle\langle\Psi^+|\nn
\\&&+(1-a-2\epsilon)|--\rangle\langle--| \}, \eea where $|\Psi^{\pm}
\rangle=\frac{1}{\sqrt{2}}(|+-\rangle\pm|-+\rangle)$ and
$\epsilon$ is an infinitesimal parameter. In the long time limit,
the concurrence of $\tilde{\rho}^{\epsilon}(\infty)$ is
$\frac{2\epsilon}{3}$ which is just the amount of singlet states
in $\tilde{\rho}^{\epsilon}$. Therefore, the singlet part of
$\tilde{\rho}^{\epsilon}$ is being preserved under evolution,
whereas the triplet states will be affected and dissipated by
decoherence. In fact, the same evidence can also be further
generalized to the initial states given by Eq.(9). Indeed, by
taking long time limit of $\rho(\infty)$, one obtain \bea
\rho(\infty)=\left(\begin{array}{cccc}
0&0&0&0\\0&\frac{S_-}{4}&-\frac{S_-}{4}&0\\0&-\frac{S_-}{4}&\frac{S_-}{4}&0\\
0&0&0&1-\frac{S_-}{2}\end{array}\right) \eea where $S_-$ is
defined earlier. The corresponding concurrence is
$C(\rho(\infty))=\frac{S_-}{2}.$ The condition with entanglement
implies $S_-\neq 0$. This result shows that, once the initial
state $\rho(0)$ contains some ingredients of singlet even with
$\rho(0)$ being no entangled state, within finite time, the state
attains stabilized entanglement. For $\rho^F$,
$C(\rho^F(\infty))=F$ with $F$ being the fraction of singlet
state. Therefore, if the amount of singlet state is nonzero,
namely $F\neq 0$, then the concurrence is also nonzero within
finite time and hence the state becomes entangled. In contrast to
the stable Bell singlet, the triplet state is unstable due to
spontaneous emission. This instability is just a consequence of
Dicke model, namely the triplet state can decay to the ground
state whereas the singlet state cannot. The reason for the singlet
state being stable is due to the vanishing of the total dipole
and, in the long wavelength limit considered here, the state
decouples from the photon bath. One should notice that, without
real photon exchange between atoms such as the case considered by
Yu and Eberly\cite{Eber} where two atoms are put in two separate
cavities, disentanglement always happens whatever the initial
state is.

To summarize, we have shown that the interactions induced by the
vacuum fluctuations provide both dipole interactions and nonlocal
photon exchanged interactions. The decoherent effect of
spontaneous emission is shown being suppressed by the nonlocal
photon exchanged interactions coming from the non-unitary
dynamics. In passing we would like to stress again that such
suppression mechanism is due to the real photon exchange process.
With the success of suppressing decoherence, it is further shown
that entanglement can be stabilized and even enhanced.
Surprisingly, the enhancement process can be saturated in finite
time. The basic condition for stabilization and enhancement of
entangled state is the existence of singlet state in any mixed
state and supporting examples are also provided in this work.

\vskip 1.0cm This work was supported by the National Science
Council of R.O.C. under the Grant No. NSC-93-2112-M-006-006 and
NSC-94-2112-M-006-016. We also thank the National Center for
Theoretical Sciences(South) of Taiwan, R.O.C. and the Center for
Quantum Information Science of National Cheng Kung University for
financial support.


\end{document}